\documentstyle[aps,epsfig,twocolumn]{revtex}
\begin{document}
\draft
\flushbottom
\twocolumn[
\hsize\textwidth\columnwidth\hsize\csname @twocolumnfalse\endcsname
\title{Surface effects 
in multiband superconductors. Application to MgB$_2$.}
\author{E. Bascones and F. Guinea \\}
\address{
Instituto de Ciencia de Materiales.
Consejo Superior de Investigaciones Cient{\'\i}ficas.
Cantoblanco. 28049 Madrid. Spain.} 
\date{\today}
\maketitle
\tightenlines
\widetext
\advance\leftskip by 57pt
\advance\rightskip by 57pt

\begin{abstract}
Metals with many bands at the Fermi level can have
different band dependent gaps in the superconducting
state. The absence of translational symmetry
at an interface can induce interband scattering
and modify the superconducting properties.
We dicuss the relevance of these effects to
recent experiments in MgB$_2$.
\end{abstract}

]
\narrowtext
\tightenlines
\section{Introduction}
Recent experiments\cite{Aetal01} report the existence
of superconductivity at nearly 40K in MgB$_2$. Its origin
is not completely elucidated. The material shows a
pronounced isotope effect\cite{Betal01}, and 
the density of states is well approximated by 
the BCS theory\cite{RSV01,Setal01}.
Tunneling experiments suggest that the superconducting 
properties at the surface of the material differ
from the expected bulk 
behavior\cite{RSV01,Setal01}. In addition, photoemission
results\cite{Tetal01} suggest the existence of an s-like gap,
$\Delta$, such that $\Delta \le 3 k_B T_c$. A possible 
explanation of this result is that the measured $\Delta$ is the
average of different gaps. The present work is motivated by
the persistent discrepancy between the gap values measured
in different experiments, and, particularly, the excellent fit
to a BCS gap too low to explain the value of the critical
temperature observed in tunneling experiments reported
in\cite{RSV01}.

Band structure calculations suggest that there are, at least,
two types of bands at the Fermi surface: a hole band, 
built up of boron $\sigma$ orbitals, with
a weak dispersion in the direction perpendicular to the
boron planes, and a broader band, built up mainly of
$\pi$ boron orbitals, which shows a significant
dispersion in the direction perpendicular to the
boron planes. Theoretical arguments favor,
as the origin of the superconductivity, the 
hole like $\sigma$ band\cite{H01,BSA01,AP01,Ketal01c}, or the $\pi$
band\cite{Ketal01b}. The existence of two bands
with different physical properties is assumed in
other models for the superconducting properties
of MgB$_2$\cite{I01,VAR01}. It has been argued that
the upper critical field can be best modelled
if the superconducting properties depend on the
specific band at the Fermi level\cite{Setal01b}.
On general grounds, it is reasonable to assume that
the $\sigma$ and $\pi$ bands in MgB$_2$
will have different contributions to the
superconducting phases, and that the superconducting gap
needs not be the same in the two bands.

The existence of many bands at the Fermi level, with
very different physical properties, is probably a
generic feature of intermetallic superconductors\cite{MRS01,Setal98}.
In these materials, it can be expected that the pairing
interaction which gives rise to the superconductivity
will depend on the details of each band.
If this is the case, there is not a uniform gap
at the Fermi level. The superconducting state
resembles, in this respect, that of an
anisotropic superconductor. The effects of interband scattering
on the bulk properties of a superconductor with two 
different bands at the Fermi level was studied
in\cite{GM97}.

Interband scattering can be induced by any defect
which breaks the translational symmetry of the
lattice, including lattice imperfections, phonons,
and surfaces or grain boundaries. Thus, in 
materials with many bands crossing the Fermi level,
one can expect the superconducting properties
near a surface to differ from those at the bulk.
This effect should be more pronounced in systems
with many irregular interfaces, such as ceramic
and granular materials.

In the present work, we study pair breaking effects
at the surfaces of
many band superconductors. The main difference with 
the theory developed in\cite{GM97} is that, when
interband scattering is restricted to the surface
region, the superconducting gaps are inhomogeneous,
because the value of the gaps within the bulk of
the system is not affected by the presence of
a surface. In order to take this effect into
account, we develop a method which differs
from the standard treatment of pair breaking 
effects in conventional\cite{AG61} or
unconventional\cite{HWE88} superconductors.
We use the simplest possible model where non trivial
effects are expected, presented in the next
section. Section III presents the main results, and
the main conclusions which can be drawn from them 
are discussed in section IV.
\section{The model}
\subsection{The hamiltonian.}
Superconductors with different bands at the Fermi level,
where each of these bands have different pairing
interactions have already been discussed in
the literature\cite{SMW59,HM91}. The simplest
model contains two bands, with two different
densities of states and pairing interactions.
We will consider the hamiltonian:
\begin{eqnarray}
{\cal H} &= &{\cal H}_0 + {\cal H}_{int} + {\cal H}_{pb}
\nonumber \\
{\cal H}_0 &= &\sum_{
i=1,2;{\bf \vec{k}},s} \epsilon_{i, {\bf \vec{k}}}
c_{i, {\bf \vec{k}},s}^\dag
c_{i, {\bf \vec{k}},s} \nonumber \\
{\cal H}_{int} &= &\sum_{
i=1,2;{\bf \vec{k}}
{\bf \vec{k}}'} - g_i c_{i, {\bf \vec{k}}\uparrow}^\dag
c_{i, - {\bf \vec{k}} \downarrow}^\dag
c_{i, {\bf \vec{k}}'\uparrow}
c_{i, - {\bf \vec{k}}' \downarrow} \nonumber \\
&- &\sum_{{\bf \vec{k}}{\bf \vec{k}}'} g'
c_{1, {\bf \vec{k}}\uparrow}^\dag
c_{1, - {\bf \vec{k}} \downarrow}^\dag
c_{2, {\bf \vec{k}}'\uparrow}
c_{2, - {\bf \vec{k}}' \downarrow} \nonumber \\
{\cal H}_{pb} &= &\sum_{s} \int d^3 {\bf \vec{r}}
V f ( z )
\psi_{1, s}^\dag ( {\bf \vec{r}} )
\psi_{2, s} ( {\bf \vec{r}} )
\label{hamil} 
\end{eqnarray}
Electrons within each band experience a different
pairing interaction, $g_i$, leading to two 
superconducting gaps. The two bands are coupled
by the interaction $g'$. Otherwise, in the absence
of interband scattering the two gaps open at
different temperatures, $T_{c,i}
\propto \omega_0 \exp{(-W_i/g_i)}$, where $W_i$ is the bandwidth,
and $\omega_0$ is a cutoff related to the pairing mechanism.
Specific heat
measurements\cite{Wetal01} seem to exclude this possibility
in MgB$_2$. 
We neglect 
intraband scattering, which does not give rise to
pair breaking effects, at least to lowest
order\cite{A59}. Finally, we assume a constant 
interband scattering term, localized near the surface.
The function $f(z)$ is peaked near the surface, located
at $z=0$. The width of $f(z)$ is of the order of
a lattice spacing.
With these restrictions, the model described
in eq.(\ref{hamil}) includes six parameters,
with dimensions of energy:
the density of states at the Fermi level of each band,
$N_i ( \epsilon_F )$, the pairing interactions, $g_i$
and $g'$,
and the scattering
potential, $V$.
\subsection{Pair breaking effects.}
\begin{figure}
\centerline{\mbox{\epsfxsize 5cm \epsfbox{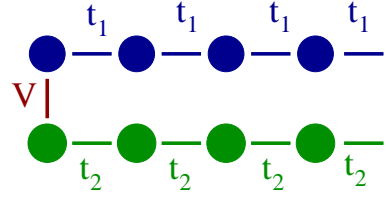}}}
\caption{Sketch of the lattice solved in the text.}
\label{sketch}
\end{figure}

The lack of translational symmetry induced by the surface
makes it convenient to solve directly the
the Bogoliubov-de Gennes equations derived
from (\ref{hamil}). We use a discrete tight binding
model for this purpose. We assume that each band can
be described by a single orbital per site, and that
there are local attractive interactions with induce
the pairing. The model reduces, in the absence of
interband scattering, to two coupled negative $U$
Hubbard models. In this basis, interband scattering
can be included by allowing for hopping from
one orbital to the other at any given lattice site.
Our discretized model in real space becomes:
\begin{eqnarray}
{\cal H} &= &{\cal H}_0 +{\cal H}_{int} + {\cal H}_{pb}
\nonumber \\
{\cal H}_0 &= &\sum_{l;ijs} t_l c^{\dag}_{l,i,s}
c_{l,j,s} + {\rm h. c.} \nonumber \\
{\cal H}_{int} &= &- \sum_{l,i} U_l c^{\dag}_{l,i,\uparrow}
c_{l,i,\downarrow} c^{\dag}_{l,i,\downarrow}
c_{l,i,\downarrow} \nonumber \\
&- &\sum_i U' c^{\dag}_{1,i,\uparrow}
c_{1,i,\downarrow} c^{\dag}_{2,i,\downarrow}
c_{2,i,\downarrow} \nonumber \\ 
{\cal H}_{pb} &= &\sum_{s,i \in i_s} V c^{\dag}_{1,i,s}
c_{2,i,s}
\label{hubbard}
\end{eqnarray} 
We assume that the lattice is a semiinfinite chain, and that
interband scattering is restricted to the outermost site,
as schematically depicted in Fig.[\ref{sketch}]. 
The BCS equations are solved using the standard
equivalence of the attractive Hubbard model in a bipartite
lattice to the repulsive Hubbard model\cite{GP95}.
The semiinfinite model is solved using transfer matrix
techniques\cite{G84}, which are described in the Appendix.
The gaps are calculated selfconsistently in a layer
of $m$ sites, which are coupled to an homogeneous chain
where the gaps take the bulk values,
where the gaps are defined as:
\begin{equation}
\Delta_i = \frac{U_i}{2} \langle c^{\dag}_{i \uparrow}
c^{\dag}_{i \downarrow} \rangle +
\frac{U'}{2} \langle c^{\dag}_{j\uparrow} 
c^{\dag}_{j\downarrow} \rangle
\end{equation}
We present results
obtained with $m = 32$. This idealized one dimensional model
can be viewed as an approximation to the inhomogeneous
layered structure 
expected near the surface.
Note that the BCS equations exclude the possibility of
one dimensional fluctuations, so that the solutions
to be discussed below do not show unphysical one
dimensional features.
\section{Results}
\begin{figure}
\centerline{\epsfig{file=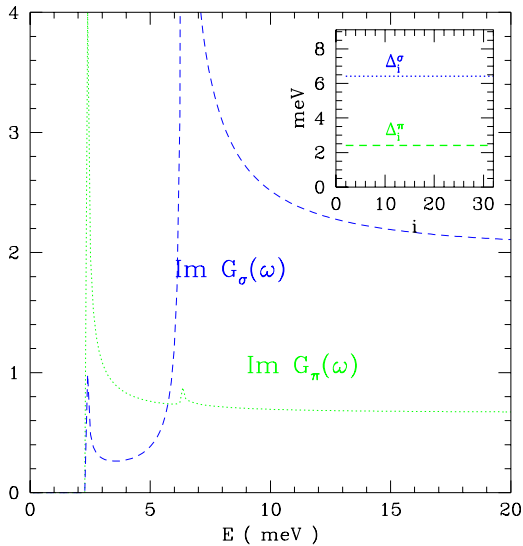,width=3in}}
\caption{Density of states at the surface at zero temperature.
The inset shows the superconducting order parameters in each
band as function of position from the surface.}
\label{fig_gap}
\end{figure}

We use as unit of energy one of the hopping parameters in
eq.(\ref{hubbard}), $t_1 = 1$. We assume that band 1 represents
the total effect of the two $\sigma$ bands in MgB$_2$.
The density of states in the $\pi$ band is roughly one third
that of the two $\sigma$ bands\cite{AP01}, and we set
$t_2 = 3 t_1$. We assume that
the superconductivity is mostly due
to the $\sigma$ band, and we choose 
the parameters $U_1 , U_2$ and $U'$ so that
the smallest gap is less than one half the largest one,
at zero temperature\cite{RSV01}. A reasonable combination is
$U_1 = 0.5 t_1 , U_2 = 0$ and $U' = 0.2 t_1$ so that
$\Delta_1 = 0.0212 t_1$ and $\Delta_2 = 0.0079 t_1$.
The critical temperature in these units is $T_c = 0.011 t_1$.
Using the experimental value $T_c \approx 40$K, the value
of the unit of energy, $t_1$ is equal to 0.3 eV, and the density
of states of the $\sigma$ band is 0.5 states/eV-cell, which
has the right order of magnitude\cite{AP01}. Note that, with
these parameters, the model is well into the BCS weak coupling
regime.
Finally, we must determine the strength of the interband
scattering at the surface, $V$. The mixing of the boron $\pi$ and
$\sigma$ bands is strongly suppressed in the bulk. This 
effect is weakened if the relative angles between nearest
neighbor B ions is modified at the surface, so that the
$\sigma$ bands are not built up of sp$^2$ orbitals only.
The interband hopping associated with these deformations
will be a fraction of the bulk hopping terms. We take in the
following $V = 0.4 t_1 \approx 0.12$eV, 
and restrict this interband hopping 
to the outermost layer.
The gaps near the surface, at zero temperature, are plotted
in the inset in Fig.[\ref{fig_gap}]. The gaps are almost uniform, suggesting
that the influence of the interband scattering on the value
of the gaps is very small.

\begin{figure}
\centerline{\epsfig{file=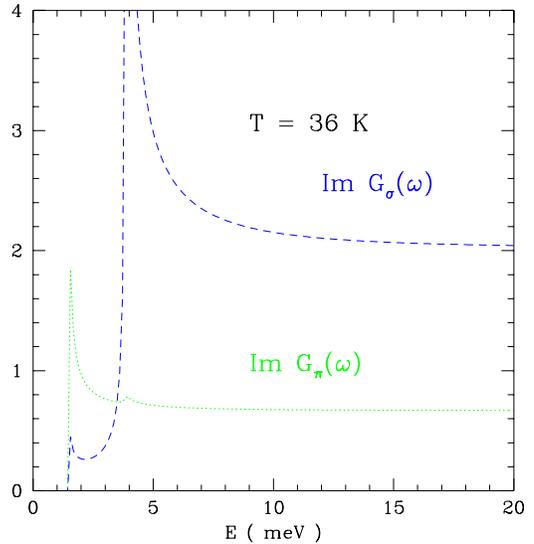,width=3in}}
\caption{Density of states at the surface at T = 36K.}
\label{fig_dost36}
\end{figure}

\begin{figure}
\centerline{\epsfig{file=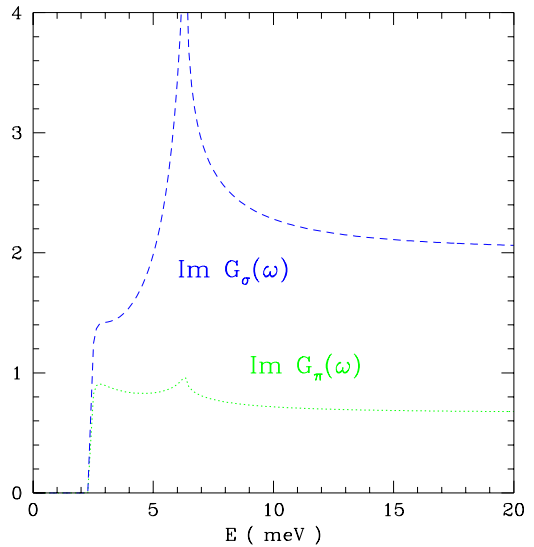,width=3in}}
\caption{Density of states at the surface at T = 0K,
in the presence of strong interband scattering at the surface.}
\label{fig_pb}
\end{figure}

Interband scattering has a stronger influence on the density of states
near the surface. The results for the outermost site are plotted
in Fig.[\ref{fig_gap}]. The coupling between the two bands
shows in the existence of peaks in the density of states 
at the two gap positions, while, in the bulk, each band
displays a single peak at the value of the corresponding gap.
The overall features in the density of states remain the same
at relatively high temperatures, as shown in
fig.[\ref{fig_dost36}], where the results at T=36K are shown.
The scale at which the smallest gap closes is determined
by the largest gap.
In some experiments, like point contact spectroscopy, the probe can 
be an additional source of interband scattering, at the position
where the measurement is being made. In the presence of strong
interband scattering at the surface, the 
perturbation in the densities of states
is more pronounced, as shown in Fig.[\ref{fig_pb}],
calculated using $V =  t_1 \approx 0.3$eV. A single smeared
gap will be observed in an experiment of this type.
It is interesting to note that, if the value of the gaps
in the bulk were of opposite sign, a midgap state,
induced by Andreev reflections, should arise.
The problem considered here maps onto that of
a dimerized chain. If the value of the dimerization
changes sign, a gap state always arises, as
extensively discussed in connexion to
solitons in polyacetilene\cite{HKSS88}.
\section{Conclusions}
We have analyzed the effects of interband scattering
at the surface, in a superconductor with a many bands
at the Fermi level, and different pairing strengths
for each band.

Surface scattering, for reasonable values of the parameters,
is inneffective in changing the gaps near the surface.
This is consistent with the fact that, in a
conventional weak coupling BCS superconductor, the
coherence length is much greater than the surface
layer where strong scattering is expected.

The influence of the surface scattering
induces significant changes in the density of states
near the surface. The density of states of each band
shows peaks at the positions of the gaps associated
with the other bands. There is, however, always a minimum
gap, below which the density of states vanishes. 
This gap, $\Delta_{min}$,
has a weak temperature dependence until close to the bulk
critical temperature.

We have used a set of parameters 
appropiate for MgB$_2$. We assume that there is a wide,
delocalized band, and a narrower and more localized band,
which determines mostly the superconducting properties. 
The wide band (derived from the
$\pi$ orbitals) has the smallest gap, and is weakly
influenced by the narrow band. The narrow band
shows stronger features at both the small and
large gaps. Tunneling experiments are highly
sensitive to the delocalization of the
wavefunction. Hence, it is possible that 
they measure the density of states
associated to the wider band ($\pi$ orbitals). 
If this is the case, the observations will
show a single gap, well approximated by the BCS
expression. This gap should have a weak
temperature dependence, until temperatures
comparable to the bulk critical temperature\cite{RSV01}.

The strength of the interband scattering at the surface studied here
can depend on the experimental setup, and it may be enhanced
in some experiments. If that is the case, a single smeared gap
will be observed, of magnitude 
comparable to 1.7 T$_{\rm c}$\cite{Ketal01a}. 

Finally, it is interesting to note that, if
the gaps in the two bands were of opposite signs,
as expected from electron-electron pairing mechanisms,
a surface state near the center of the gap should
appear.
\section{Acknowledgements}
We are thankful to S. Vieira, H. Suderow and G. Rubio-Bollinger
for many helpful discussions. Financial support
from CICyT (Spain) through grant PB96/0875 is gratefully
acknowledged. 

\section{Appendix}
The problem defined by the hamiltonian in eq.(\ref{hubbard})
with the geometry shown in Fig.[\ref{sketch}] reduces to the
calculation of the density of states in $2 m$ sites, with
gap values $\Delta_i$ and connected by hopping terms
which can be either $t_1 , t_2$ or $V$. The attractive Hubbard
model, $U_i < 0$, can be mapped onto the repulsive Hubbard model,
in a bipartite lattice,
by the transformation:
\begin{eqnarray}
c^{\dag}_{i \uparrow} &\rightarrow &d^{\dag}_{i \uparrow}
\nonumber \\
c_{i \uparrow} &\rightarrow &d_{i \uparrow}
\nonumber \\
c^{\dag}_{i \downarrow} &\rightarrow &
(-1)^i d_{i \downarrow}
\nonumber \\
c_{i \downarrow} &\rightarrow &
(-1)^i d^{\dag}_{i \downarrow}
\end{eqnarray}
The superconducting order parameter,
$\Psi_i = \langle c^{\dag}_{i \uparrow} c^{\dag}_{i \downarrow}
\rangle$, is mapped onto a staggered magnetization in the transverse
direction. At half filling, there is an additional symmetry which allows
us to rotate this magnetization to the z-axis. Then,
$\Psi_i = 
\langle (-1)^i ( c^{\dag}_{i \uparrow} c_{i \uparrow} -
c^{\dag}_{i \downarrow} c_{i \downarrow} ) \rangle$.
In this representation, the hamiltonian does not mix the spins,
and it can be decomposed into two boxes, one for each spin
direction. The problem is reduced to the calculation of
the density of states in a tight binding chain with 
variable hoppings, $t_{i,i\pm 1}$, and energy levels,
$\epsilon_i$, which are related to the local value
of the gaps. The gaps must be determined selfconsistency
from the values of the $\Psi_i$'s.

The fractions:
\begin{equation}
g_{n,n \pm 1} ( \omega ) = 
t_{n,n \pm 1} \frac{G_{n,n'} ( \omega )}
{G_{n \pm 1 , n'} ( \omega )}
\end{equation}
are independent of $n'$, and satisfy:
\begin{equation}
g_{n,n \pm 1} ( \omega ) = \frac{t_{n,n \pm 1}^2}{
\omega - \epsilon_{n,n \pm 1} - t_{n \pm 1 , n \pm 2}
g_{n \pm 1 , n \pm 2} ( \omega )}
\end{equation}
and:
\begin{equation}
G_{n,n} ( \omega )
= \frac{1}{\omega - \epsilon_n - t_{n,n-1} g_{n,n-1} ( \omega )
- t_{n,n+1} g_{n,n+1} ( \omega )}
\end{equation}
Thus, the problem can be solved by iteration from the boundaries,
provided that one knows the values of $g_{\pm m , \pm m \pm 1}$.
These values can be easily be calculated, if one assumes that
the values of the $\epsilon_i$'s are constant beyond  
position $m$\cite{FY75}. Finally, the selfconsistency requirement
for the values of $\epsilon_i , i=1,2m$ must be satisfied.

\end{document}